\documentclass[12pt,a4paper]{ijicic}

\def\currentvolume{8}
\def\currentissue{6}
\def\currentyear{2012}
\def\currentmonth{January}
\def\ppages{1--17}

\usepackage[dvips]{graphicx}
\usepackage{epsfig}
\usepackage{epsf}
\usepackage{ cite, url, subfigure, longtable, enumerate, setspace,amsmath, listings, color}

\title[Traffic Engineering on a Novel resource reservation over Mobile IPv6]
      {Traffic Engineering Based on Effective Envelope Algorithm on Novel Resource Reservation Method over Mobile Internet Protocol Version 6}

\author[R. Malekian and A. H. Abdullah]{}

\begin{document}

\maketitle

\centerline{\scshape  Reza Malekian$^1$ and  Abdul Hanan Abdullah$^1$}
 \medskip
{\footnotesize
\centerline{$^1$Faculty of Computer Science and Information Systems}
\centerline{Universiti Teknologi Malaysia}
\centerline{81300 Skudai, Johor, Malaysia}
\centerline{mreza7@live.utm.my} }
\medskip

\centerline{Received May 2011; revised September 2011}


\medskip

\begin{abstract}

{\em The first decade of the 21st century has seen tremendous improvements in mobile internet and its technologies. The high traffic volume of services such as video conference and other real-time traffic applications are imposing a great challenge on networks. In the meantime, demand for the use of mobile devices in computation and communication such as smart phones, personal digital assistants, and mobile-enabled laptops has grown rapidly. These services have driven the demand for increasing and guaranteing bandwidth requirements in the network. A direction of this paper is in the case of resource reservation protocol (RSVP) over mobile IPv6 networks. There are numbers of proposed solutions for RSVP and quality of service provision over mobile IPv6 networks, but most of them using advanced resource reservation. In this paper, we propose a mathematical model to determine maximum end-to-end delay bound through intermediate routers along the network. These bounds are sent back to the home agent for further processing. Once the home agent receives maximum end-to-end delay bounds, it calculates cumulative bound and compares this bound with the desired application end-to-end delay bound to make final decision on resource reservation. This approach improves network resource utilization. }\\
{\bf Keywords:} Mobile IPv6, Resource reservation, Traffic engineering, Effective envelop, Resource utilization.

\end{abstract}

\end{document}